\documentclass{svproc}
\usepackage{url}

\usepackage{graphicx}
\usepackage{IEEEtrantools}
\usepackage{subfigure}

\usepackage{floatrow}

\begin{document}
\mainmatter              
\title{The Unfulfilled Potential of Data-Driven \\Decision Making in Agile Software Development}
\titlerunning{The Unfulfilled Potential of Data-Driven Decision Making in Agile}  
%
\author{Richard Berntsson Svensson \and Robert Feldt \and Richard Torkar}
\authorrunning{Richard Berntsson Svensson \and Robert Feldt \and Richard Torkar} 
%
\tocauthor{Richard Berntsson Svensson and Robert Feldt and Richard Torkar}
\institute{Department of Computer Science and Engineering, Chalmers $|$ University of Gothenburg, Gothenburg, Sweden\\
\email{richard.berntsson.svensson@gu.se, robert.feldt@chalmers.se, richard.torkar@cse.gu.se}}

\maketitle              

\begin{abstract}
With the general trend towards data-driven decision making (DDDM), organizations are looking for ways to use DDDM to improve their decisions. However, few studies have looked into the practitioners view of DDDM, in particular for agile organizations. In this paper we investigated the experiences of using DDDM, and how data can improve decision making. An emailed questionnaire was sent out to 124 industry practitioners in agile software developing companies, of which 84 answered. The results show that few practitioners indicated a wide-spread use of DDDM in their current decision making practices. The practitioners were more positive to its future use for higher-level and more general decision making, fairly positive to its use for requirements elicitation and prioritization decisions, while being less positive to its future use at the team level. The practitioners do see a lot of potential for DDDM in an agile context; however, currently unfulfilled.
\end{abstract}
\section{Introduction}
When developing software-intensive products, agile methods have become the \emph{de facto} way to develop software across almost every industry. The introduction of agile methodologies has changed the way software is developed~\cite{Petersen2009}, how Requirements Engineering (RE) is conducted~\cite{Schon2017}, and how decisions are made~\cite{Moe2012}. In transitioning to Agile Software Development (ASD), learning about the customers, collecting customer\slash user feedback, and involving a customer representative in development, requirements engineering, and decision making, are important~\cite{Olsson2014}. In addition, ASD teams, due to delivering working software in short iterations, are frequently involved in short-term decisions and need to adopt to a fast decision making process~\cite{Cockburn2001}.

With digital networks connecting an increasing number of people, devices, and products, a vast amount of diverse data is available. Industries gather data and knowledge from their customers, suppliers, alliance partners, and competitors. For example, mobile phones, cars, transportation vehicles, and automation systems, are developed to generate data about their customers and usage of their activities. This diverse data is not only generated internally within software-intensive companies, but also from public, proprietary, and purchased sources~\cite{Provost2013}. Software developing companies need to focus on exploiting the available data to gain competitive advantages~\cite{Provost2013}, which will transform how business are generated, how RE is performed, and how decisions are made~\cite{Walid2016}. In particular, the recent resurgence of interest in artificial intelligence (AI) and machine learning (ML) accelerates these trends due to their promise of more automated and powerful data analysis.

However, despite the vast amount of data that is available for decision making, the decisions and selection of what to include in the next product release cycle, are commonly based on the product managements and\slash or stakeholders' previous experiences, opinions, intuitions, various criteria, arguments, or a combination of one or several of these information sources~\cite{Olsson2014,Walid2016}. These decisions are typically subjective, frequently inconsistent, and often lack explanations as well as links to which data and evidence they were based on. Moreover, when stakeholders make decisions based on, e.g., opinions, intuitions, and arguments, the decisions are more likely to be influenced by politics and individual agendas~\cite{Milne2012,magazinius2012investigating,magazinius2011confirming} rather than, e.g., business opportunities or customer value. In addition, even when data is more clearly being taken into account in decisions, too much data and information may distract the decision maker rather then inform them. According to Wnuk et al.~\cite{Wnuk2013}, irrelevant information is visible in practitioner backlogs to a large extent today, and recent research shows that it can negatively impact decisions~\cite{Green2017}.

In order to benefit from data-driven decision making (DDDM), not only is the quality of the processing techniques and tools directly related to the quality of the decisions~\cite{Janssen2017}, but also the quality of the visualizations used to support decision makers~\cite{Janssen2017}. While visualization of software engineering data has shown promise in supporting practitioners' decisions, the focus has often been on specific phases or problems, e.g., testing and quality assurance~\cite{feldt2013supporting}, rather than throughout development processes and in agile settings. In the literature, most of the attention in DDDM has focused on the development of new techniques, technologies, and tools for data processing~\cite{Chen2014}, while few (if any) have investigated DDDM from the practitioners' perspectives and the specific and important context of agile development has not been in focus.

This paper presents the results of an empirical study that includes data collected through an emailed questionnaire with 84 respondents from 28 agile software developing companies from 9 domains. The study investigate how common the use of data for decision making is in industry today, how often data is used, the respondents opinions about the usage of data in the future, and how data can improve decision making.

The remainder of this paper is organized as follows. In Section~\ref{RW}, we outline the background to data-driven decision making. Section~\ref{RM} describes the research methodology, while Section~\ref{Analysis} presents an overall statistical analysis of the data. Section~\ref{ResandDisc} presents and discuss the results, and finally Section~\ref{Con} presents the conclusions.

\section{Background}\label{RW}
Data-driven decision making (DDDM) has become a critical ability for organizational success. Several studies have demonstrated the benefits of DDDM, e.g., Brynjolfsson et al.~\cite{Brynjolfsson2011} showed that DDDM is strongly related to higher productivity, higher return on assets, return on equity, and market value.

In the literature, there are several defined steps in DDDM, starting with data capturing and resulting in decision making. For example, Chen and Zhang~\cite{Chen2014} identify five steps; data recording, data cleaning\slash integration\slash representation, data analysis, data visualization\slash interpretation, and decision making. Although steps are identified, most of the attention in the literature has focused on the development of new techniques, technologies, and tools. Techniques for DDDM involve a number of disciplines with a number of specific techniques and tools in each discipline. For example, fundamental mathematics, statistics, and optimization tools are used as input to data analysis techniques such as data mining, machine learning, neural networks, signal processing, and visualization methods~\cite{Chen2014}. Current DDDM tools can be divided into three categories: batch processing tools, stream processing tools, and interactive analysis tools~\cite{Chen2014}. For more details about different techniques, technologies, and tools, we refer to~\cite{Chen2014}. We also see an increased interest in applying AI and machine learning in a software engineering context \cite{feldt2018ways} and supporting decisions during development is one of the key application types.

The quality of the decisions when using DDDM may improve or degrade based on the quality of the data and the processing techniques and tools~\cite{Janssen2017}. However, the quality of the decisions are not only based on pre-processing techniques, processing techniques and tools, it is also related to the quality of the visualizations of the data to the decision makers, the decision makers' understanding and knowledge about the data sources, the decision makers' ability to interpret data processed data, and the decision makers' knowledge about the relationships of the data~\cite{Janssen2017}. As one example, Feldt et al.~\cite{feldt2013supporting} showed how visualisation of testing-related data, without any advanced modeling, could foster understanding and support decisions around software quality in an iterative development context. Thus, in order to benefit from DDDM, it is important to focus also on other aspects than just the pre-processing and processing techniques, technologies, and tools.

\section{Research method}\label{RM}
The objective of this study was to investigate how common the use of data for decision making is in industry today, how often data is used, and the respondents' opinions about the usage of data in the future, with a special focus on the agile context in which modern-day software is developed. Given the objective, and that the research questions are geared towards the opinions of the respondents, we chose to use a survey as the research method and emailed a questionnaire for data collection. Surveys are an appropriate strategy for getting empirical descriptions about trends, attitude and\slash or opinions of the studied population~\cite{Punter2003,Robson2002}. In addition, surveys are useful for analyzing large populations, given an adequate response rate~\cite{Creswell2011,Gliner2000}. The motivation for using an emailed questionnaire was to maximize coverage and participation. The following research questions provided the focus for the empirical investigation:

\begin{itemize}
\item \textbf{RQ1:} How do software practitioners view data as part of decision making in agile software developing companies?
\item \textbf{RQ2:} To what extent is data used for decision making and requirements engineering in agile software developing companies?
\item \textbf{RQ3:} How can data be used to improve future decisions in agile software developing companies?
\end{itemize}

\subsection{Survey study}
The survey was executed through the creation of an emailed questionnaire that was designed based on the research questions using a mix of open-ended and closed questions~\cite{Robson2002}. In order to test the reliability and validity of the survey instrument, a pilot study was conducted with one industry practitioner. Based on the feedback from the pilot study, the survey instrument was (lightly) revised. The instrument (see Table \ref{tab:instrument}) had three parts. The first part gathered demographic information about the respondents. The second part mainly addressed how, and how often data is used in decision making today, while the third part focused mainly on how data can be used for decision making in the future. Part 1 only contained free-text questions. All of the questions in Parts 2 and 3 contained Likert-type scale and free-text questions. The free-text area was added to allow the respondents to expand and\slash or explain their answer.

\begin{table}
  \floatsetup{}
  \begin{floatrow}[1]
    \ttabbox%
    {\begin{tabular}{ |l | p{11cm}| }
    \hline
    \textbf{ID} & \textbf{Question}  \\ 
 \hline
      Q0 & What company do you work for?, How many employees does your company have?,  What role do you generally have in your work?, What software development process do you use?\\
 Q1 & Data is important for decision-making  \\
 Q2& Data is highly valued for decision-making  \\ 
 Q3& Data is treated as an asset  \\
 Q4& Data is used to identify new business opportunities  \\
 Q5& Data is used to predict future trends and behavior  \\
 Q6& Decision makers use data for decision-making  \\
 Q7& Teams use data for decision-making  \\
 Q8& Data is used as part of requirements elicitation/identification  \\
 Q9& Data is used for prioritization of requirements/features  \\
 Q10& Data should be important for decision-making  \\
 Q11& Data should be highly valued for decision-making  \\
 Q12& Data should be treated as an asset  \\
 Q13& Data should be used to identify new business opportunities  \\
 Q14& Data should be used to predict future trends and behavior  \\
 Q15& Decision makers should use  data for decision-making  \\
 Q16& Teams should use data for decision-making  \\
 Q17& Data should be used as part of requirements elicitation/identification  \\
 Q18& Data should be used for prioritization of requirements/features  \\
      \hline
      \end{tabular}}
    {\caption{Survey instrument}
      \label{tab:instrument}}
\end{floatrow}
\end{table}

\paragraph{Data collection.} Subjects were sampled primarily through personal contacts and previous collaborators in industry and we encouraged them to also spread the survey within their organisations.
Hence, the sample can be described as convenience sampling~\cite{Robson2002}. We provided the contacts with the questionnaire (emailed questionnaires) and information about the goals of the survey, and asked them to answer the questions and to spread the questionnaire to their colleagues. Each contact person reported back how many people they had forwarded the questionnaire to. A total of 124 subjects received the questionnaire, and 84 completed the mandatory questions and returned the questionnaire to the researchers. That is, we obtained a response rate of 67.7\%. Without going through personal contacts in industry we likely would not have been able to get this high a response rate.

\paragraph{Data analysis.} The data was analyzed using descriptive statistics with diverging stacked bar charts for the graphical visualization. In addition, we built a linear model (ordered logit) using a Bayesian approach~\cite{gelman2013bayesian,stan} to statistically analyse the data. The analysis is described in more detail in Section~\ref{Analysis}.

\subsection{Validity threats}
To avoid evaluation apprehension (\textit{construct validity})~\cite{Wohlin2000}, we guaranteed the respondents complete anonymity. Another threat is `hypothesis guessing'~\cite{Wohlin2000}, which was minimized by clearly expressing the need for honesty in the instructions to the respondents; however, it is not possible to completely dismiss this threat. In addition, the background of the subjects, e.g., experience, may influence the results; however, since the respondents have different competences and roles we believe that this risk is limited. It is not possible to exclude the possibility that the respondents misunderstood the questions (\textit{conclusion validity})~\cite{Wohlin2000}. To minimize this threat, we conducted a pilot study with an industry practitioner, which also minimized the threat of instrumentation (\textit{internal validity})~\cite{Wohlin2000}. One threat that cannot be ignored is the interest of the respondents in the topic, which may influence the representativeness. This is difficult to counter since the willingness to participate and the interest in the topic may be linked. There are also threats to validity based on selection bias and the convenience sampling; even though we sent to most of our contacts in agile software organisations and approached them in a standardised way, the final sample might not be representative for a global population of developers. For example, they were all from organisations in Sweden.

\section{Analysis}\label{Analysis}
\begin{figure*}
    \centering
    \begin{subfigure}
        \centering
        \includegraphics[scale=0.38]{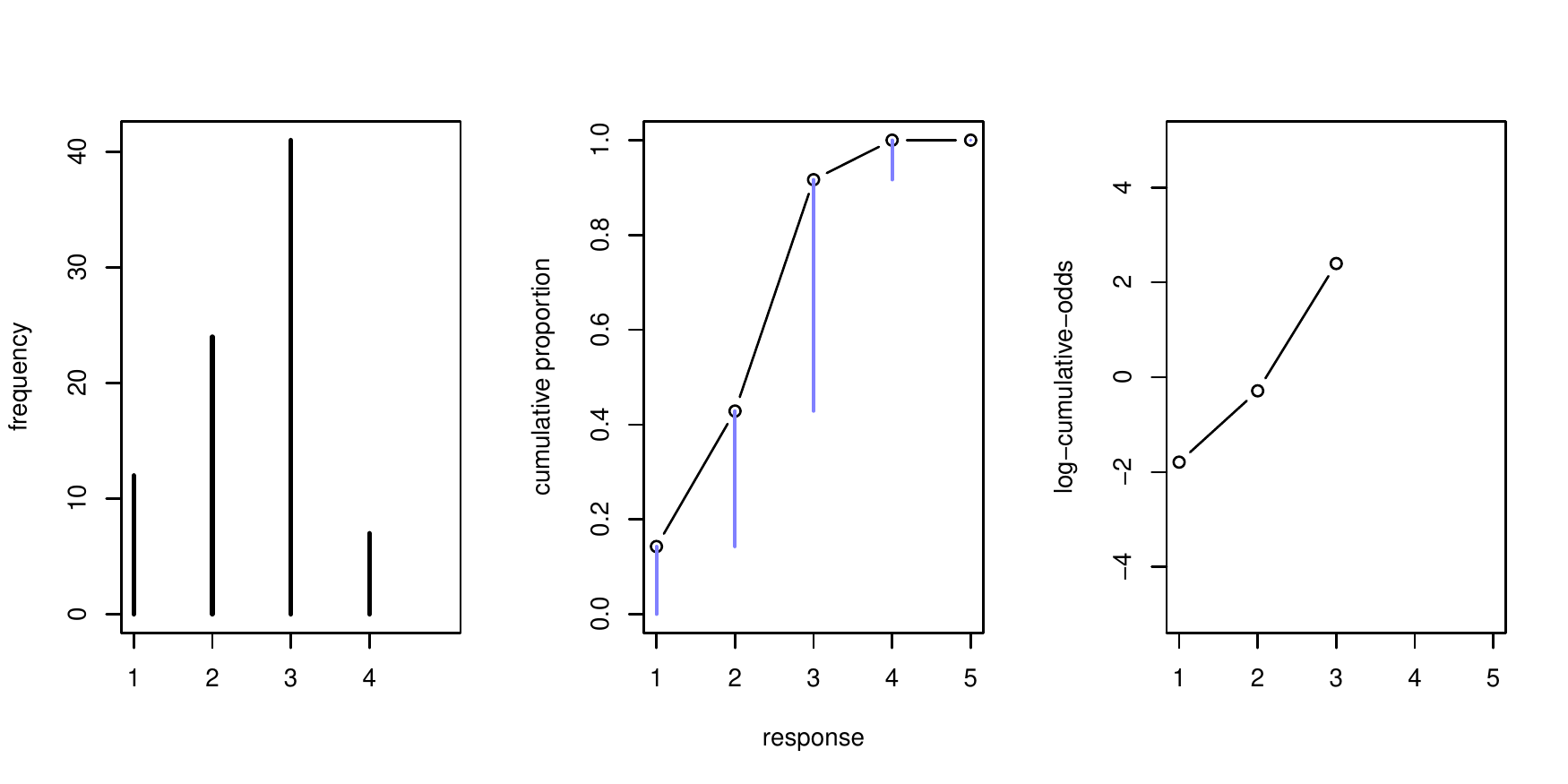}
        
    \end{subfigure}
    
    \begin{subfigure}
        \centering
        \includegraphics[scale=0.38]{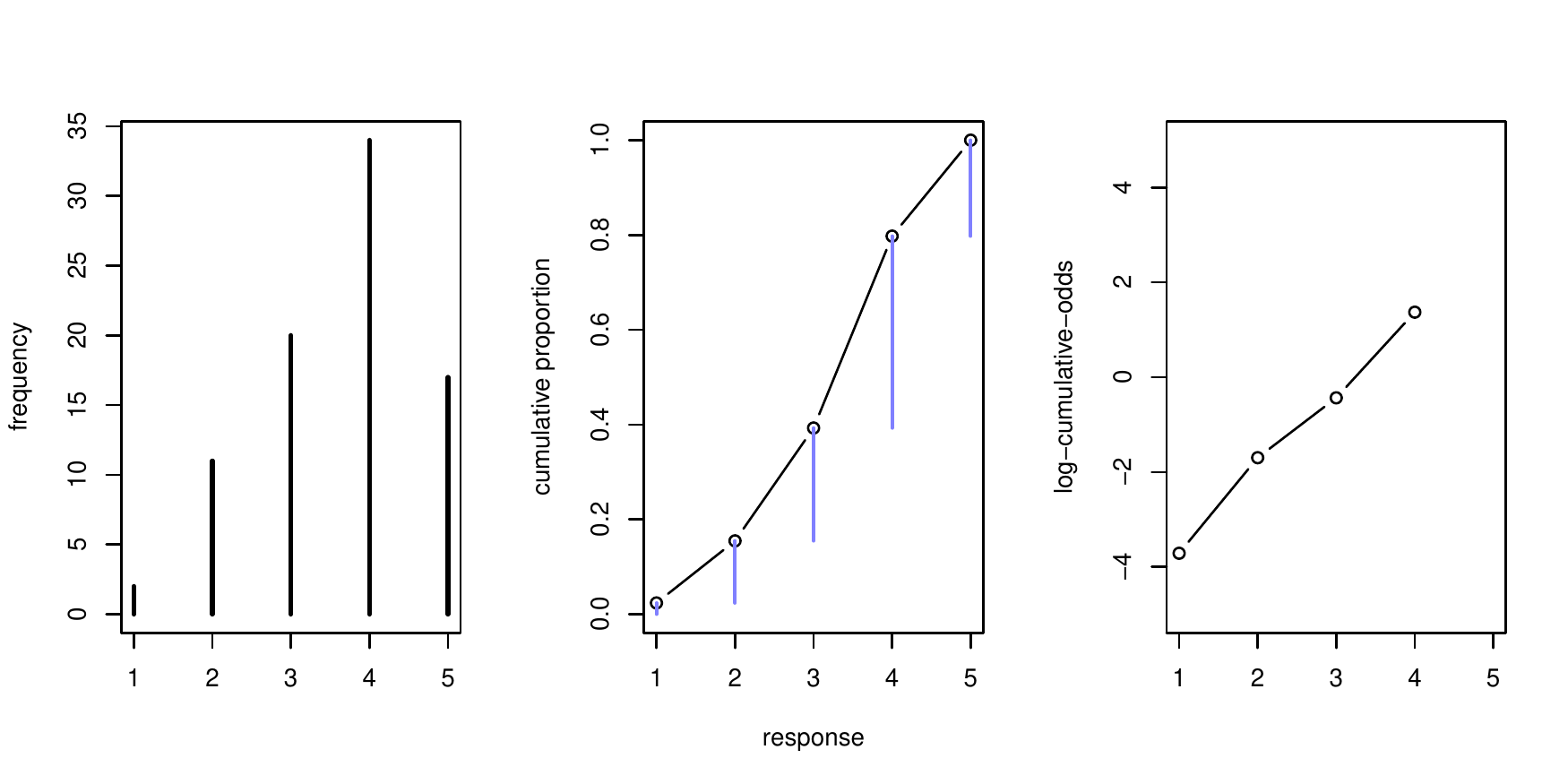}
    
    \end{subfigure}
    \caption{Plots of Q16 (top) and Q17 (bottom). Left: Histogram of discrete response in the sample. Middle: Cumulative proportion of each response. Right: Logarithm of cumulative odds of each response. Note that the log-cumulative-odds of Level 5 is infinity if there are responses among all five levels $[1, \ldots, 5]$, as in Q17 (for Q16 there were no responses on Level 5).}\label{fig:questions}
\end{figure*}

To plot and assess visually the difference between distributions of responses in Likert scale data is hard. As an example, if we examine Figure~\ref{fig:questions}, we see that there is a difference between the distribution of answers on two questions (Q16, on top in the figure, and Q17, on bottom) but it is not clear how to judge how large the difference is. Also, if we only use descriptive statistics, which is the default analysis technique for survey data in software engineering, it is difficult to assess the uncertainty of our conclusions. In contrast, a Bayesian statistical analysis does not have the same problem. Thus, in line with recent arguments for use of Bayesian methods in empirical software engineering we thus, first, start with such an analysis~\cite{furia2018bayesian,torkar2018arguing}.

In order to assess differences in Likert scale data one could assume normality and use a $t$-test, or make use of some of the non-parametric tests such as Mann-Whitney $U$ or $\chi^2$. 

However, Likert scale data is not only categorical, it is also of an ordered nature but where we cannot assume that the `distance' between consecutive pairs of answers is the same. Thus it is not clear that we can assume the data is normally distributed or that the distribution of scores for different answers has the same shape (distribution family) \cite{burknercharpentier2018}. Given these problems, in our view, the most conservative approach to analyze Likert scale data is to build a simple linear model using a Bayesian approach but keeping data categorical~\cite{gelman2013bayesian,stan}. This way we will get a posterior distribution with which we can assess uncertainty. To this end we build two overall models to study the general trends in our data:

{\scriptsize
\begin{IEEEeqnarray}{rCl}
R_i & \sim & \mathrm{Ordered}(p_i)\nonumber\\
\mathrm{logit}(p_i) & = & \beta_T * \mathrm{temporal}_i\nonumber\\
\beta_T & \sim & \mathcal{N}(0, 10)
\label{eq:1}
\end{IEEEeqnarray}
}

\setlength\abovedisplayskip{0pt}

{\scriptsize
\begin{IEEEeqnarray}{rCl}
R_i & \sim & \mathrm{Ordered}(p_i)\nonumber\\
\mathrm{logit}(p_i) & = & \beta_Q * \mathrm{req}_i\nonumber\\
\beta_Q & \sim & \mathcal{N}(0, 10)
\label{eq:2}
\end{IEEEeqnarray}
}

where $R_i$ is the $i$th response with an ordered categorical outcome, and Model 1 (Eq.~\ref{eq:1}) compares the answers for questions about the present (Questions 1--9, see Figure~\ref{fig:viewofdata}) versus future (Questions 10--18, see Figure~\ref{fig:viewofdata}) use while Model 2 (Eq.~\ref{eq:2}) compares the non-RE (Questions 13--16, see Figure~\ref{fig:useofdata}) versus the RE-specific (Questions 17--18, see Figure~\ref{fig:useofdata}) questions. We use the $\mathrm{logit}$ link function to translate the linear model's real numbers to probability mass (and hence constrain it to lie between zero and one). The linear model (in Eq.~\ref{eq:1}) then is simply a parameter $\beta_T$ that we will estimate given the data at hand (temporal). The data is coded as 0\slash 1, representing `present' (today) and `future', respectively. Finally, we assign a prior to $\beta_T$, $\mathcal{N}(0, 10)$, with a mean of 0 and a large variance of 10. This is a (very) weakly informative prior that only gives a pressure towards realistic parameter values. We also verified that the analysis was not sensitive to the prior selection (i.e., a sensitivity analysis was conducted).

For the other model (Eq.~\ref{eq:2}) we simply change the parameter. Instead of estimating $\beta_T$ using `temporal' data, we estimate $\beta_Q$ for our variable `question', which is coded 0\slash 1, representing question with a `non-RE' (Q13--16) and `RE' focus (Q17--18), respectively.\footnote{The models overall sampled well with mixed chains, $\hat{R} \ll 1.1$, and an effective sample size of $\mathrm{n}_\mathrm{eff} \gg 0.2$.}

\begin{figure*}
    \centering
    \includegraphics[scale=0.25]{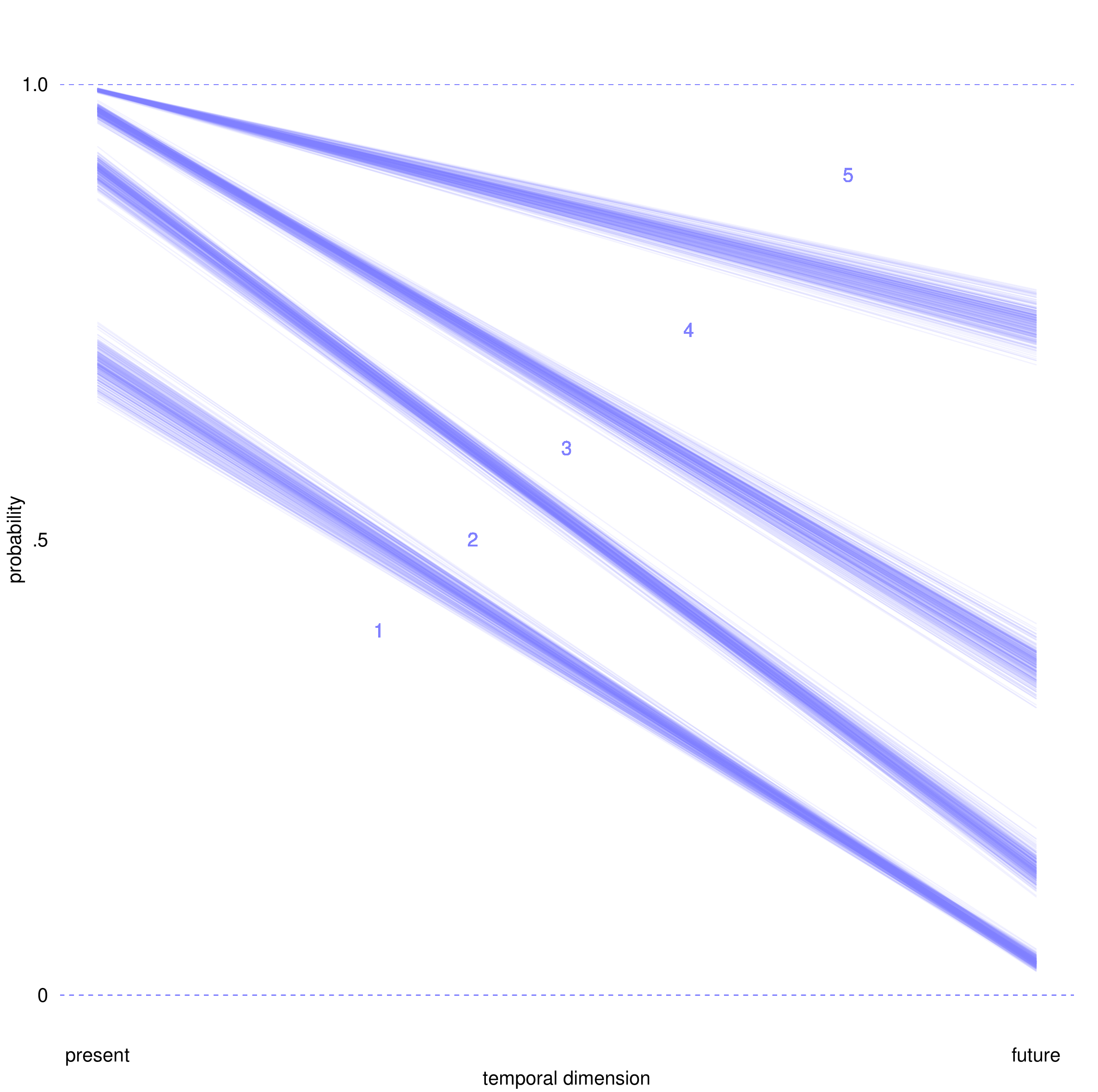}
    \caption{Posterior predictions (250 draws) of the ordered categorical model (present vs.\ future perspective). As is clearly evident, the probability for lower Likert scale values, e.g., 1 or 2, is much higher when the perspective is `present', compared to `future', i.e., everything is shifted upwards. This indicates less agreement at present and more agreement for the future, i.e. there is unfulfilled potential since the present state has a higher percentage of low disagreement answers.}
    \label{fig:temp}
\end{figure*}

Figure~\ref{fig:temp} visualizes the results from running the first model and drawing 250 samples from the posterior distribution. It is obvious that low Likert scale values are much more common for the `present' compared to the `future' category. For example, we see that the number of answers of option 1 (`Strongly disagree') is roughly around 70\% for questions about the present (today) state but decreases down to only 5\% for the future state. We can also see that the uncertainty is not large with variations only in the range of 1--7\% for all the answer alternatives.

\begin{figure*}
    \centering
    \includegraphics[scale=0.25]{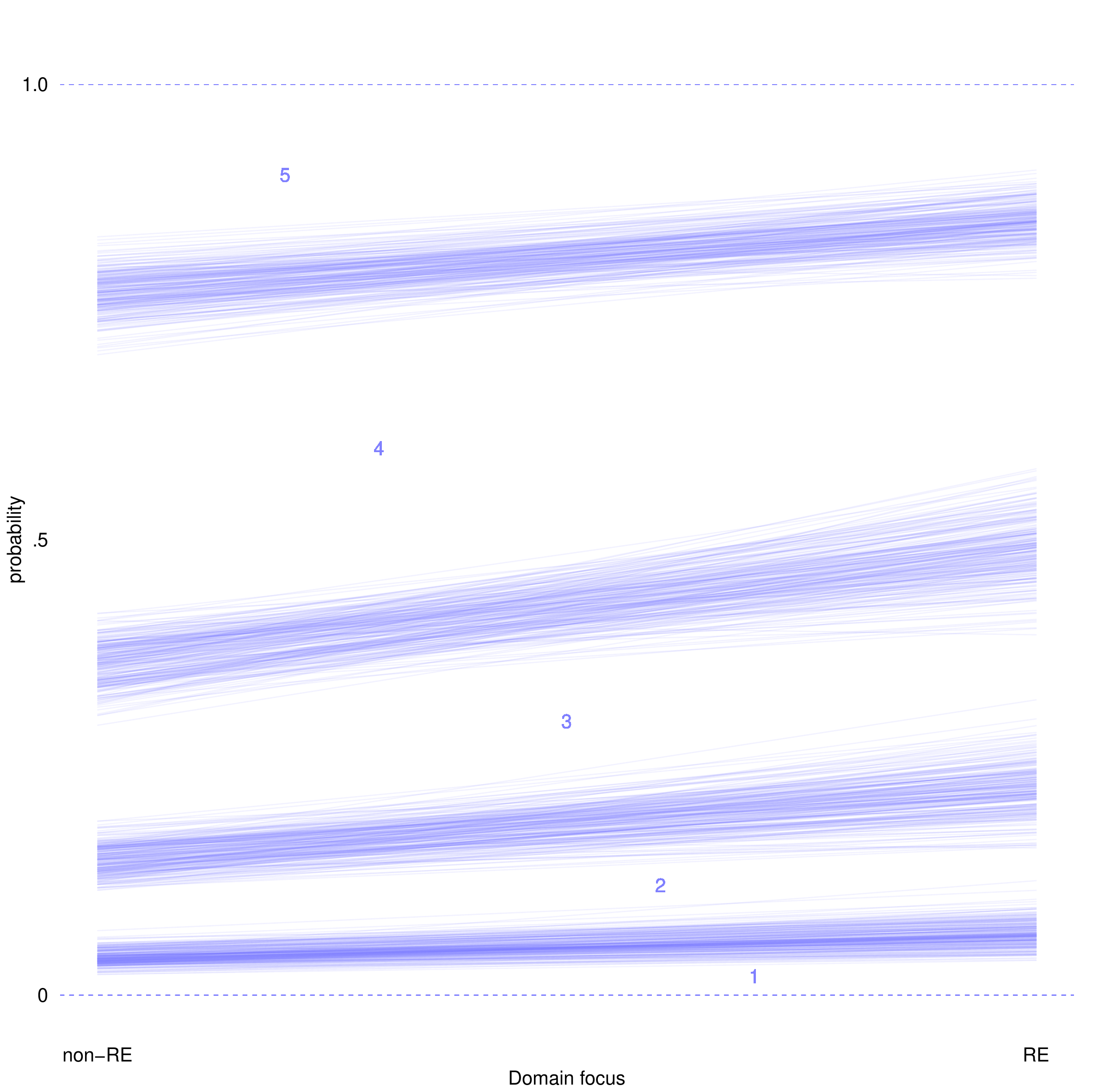}
    \caption{Posterior predictions (250 draws) of the ordered categorical model (non-RE vs.\ RE perspective). We see some trends, i.e., respondents are more positive in non-RE focused questions, but there is quite much uncertainty visualized here by the broader clusters of lines. However, the parameter $\beta_Q$, which represents the domain focus, indicates that the trend is non-negligible ($\mu=-0.53$ $\mathrm{HPDI}_{95\%}[-0.87, -0.19]$).}
    \label{fig:re}
\end{figure*}

When comparing non-RE and RE questions using Model 2 in Figure~\ref{fig:re}, we can also see some trends even if they are less clear and the uncertainty is higher as visualized by the, relatively speaking, broader bands of posterior predictions. However, the model clearly shows that we see a difference between non-RE and RE related questions with the average of the $\beta_Q$, being $\mu=-0.53$ $\mathrm{HPDI}_{95\%}[-0.87, -0.19]$, i.e., the 95\% highest posterior density estimate (HPDI) does not cross 0. This indicates that answers to the RE questions are generally lower (i.e. towards more disagreement with the statement in the question) than for the non-RE ones and that this difference is clear.

After this detailed, statistical analysis of the general trends in the responses the following Section will discuss the results in more detail.

\section{Results and Discussion}\label{ResandDisc}
This section presents the results of the survey, organized according to the research questions in Section \ref{RM}.

\subsection{Survey respondent demographics}
A total of 84 industry practitioners completed the questions of the survey. The respondents come from 28 agile software developing companies varying in size and domain. In total, the respondents came from nine different domains, with the top three being Telecommunication (27\%), Consulting (18\%), and Transportation (13\%), see Table~\ref{DemoDomains}. The size of the companies where the respondents work, in terms of number of employees, ranges from 25 up to 5,000. With respect to the respondents' roles, see Table~\ref{DemoRoles}, the top three are developers (17\%), scrum masters (15\%), and product owners (14\%) with a fairly even distribution of other, common roles also represented. For the development processes used at the companies see Table~\ref{DemoDevProc} where Scrum (43\%) is the most used, followed by (the general option) Agile (29\%), Kanban (15\%), and then DevOps (12\%). Note that the Agile category means that a respondent did not specify which agile methodology they used. Overall, we consider these respondents representative for a broad set of domains, roles and sizes of companies, even if they are all active in a Swedish context. The one role that is less clearly represented is Requirements Engineer although several of the respondents also partly do work with requirements in one form or another, as is common in agile development.

\begin{table}
  \floatsetup{}
  \begin{floatrow}[2]
    \ttabbox%
    {\begin{tabular}{ p{3cm} | l }
      \hline
      \textbf{Domain} & \textbf{Respondents} \\
      \hline
      Telecommunication & 27\% \\ 
      Consulting & 18\% \\ 
      Transportation & 13\% \\ 
      Consumer electronics & 11\% \\ 
      Surveillance & 10\% \\ 
      Control systems & 8\% \\ 
      Retail & 5\% \\ 
      Camera & 5\% \\ 
      Banking & 4\% \\
      \hline
      \end{tabular}}
    {\caption{Distribution of respondents based on domains}
      \label{DemoDomains}}
    \hfill%
    \ttabbox%
    {\begin{tabular}{ p{3.5cm} | l }
      \hline
      \textbf{Roles} & \textbf{Respondents}\\
      \hline
      Developer & 17\% \\ 
      Scrum master & 15\% \\ 
      Product Owner & 14\% \\ 
      Project manager & 11\% \\ 
      Tester & 11\% \\ 
      Senior software engineer & 11\% \\ 
      Product manager & 10\% \\ 
      Architect & 6\% \\ 
      Requirements engineer & 6\% \\
      \hline
      \end{tabular}}
    {\caption{Distribution of respondents based on roles}
      \label{DemoRoles}}
  \end{floatrow}
  \vspace*{1cm}
  \begin{floatrow}[2]
    \ttabbox%
    {\begin{tabular}{ p{4cm} | l }
      \hline
      \textbf{Development process} & \textbf{Respondents}\\
      \hline
      Scrum & 43\% \\ 
      Agile & 29\% \\ 
      Kanban & 15\% \\ 
      DevOps & 12\% \\
      XP & 1\% \\
      \hline
      \end{tabular}}
    {\caption{Distribution of respondents based on development process}
      \label{DemoDevProc}}
    \hfill%
  \end{floatrow}
\end{table}%

\subsection{View of data in decision making (RQ1)}
\begin{figure}
    \centering
    \includegraphics[scale=0.6]{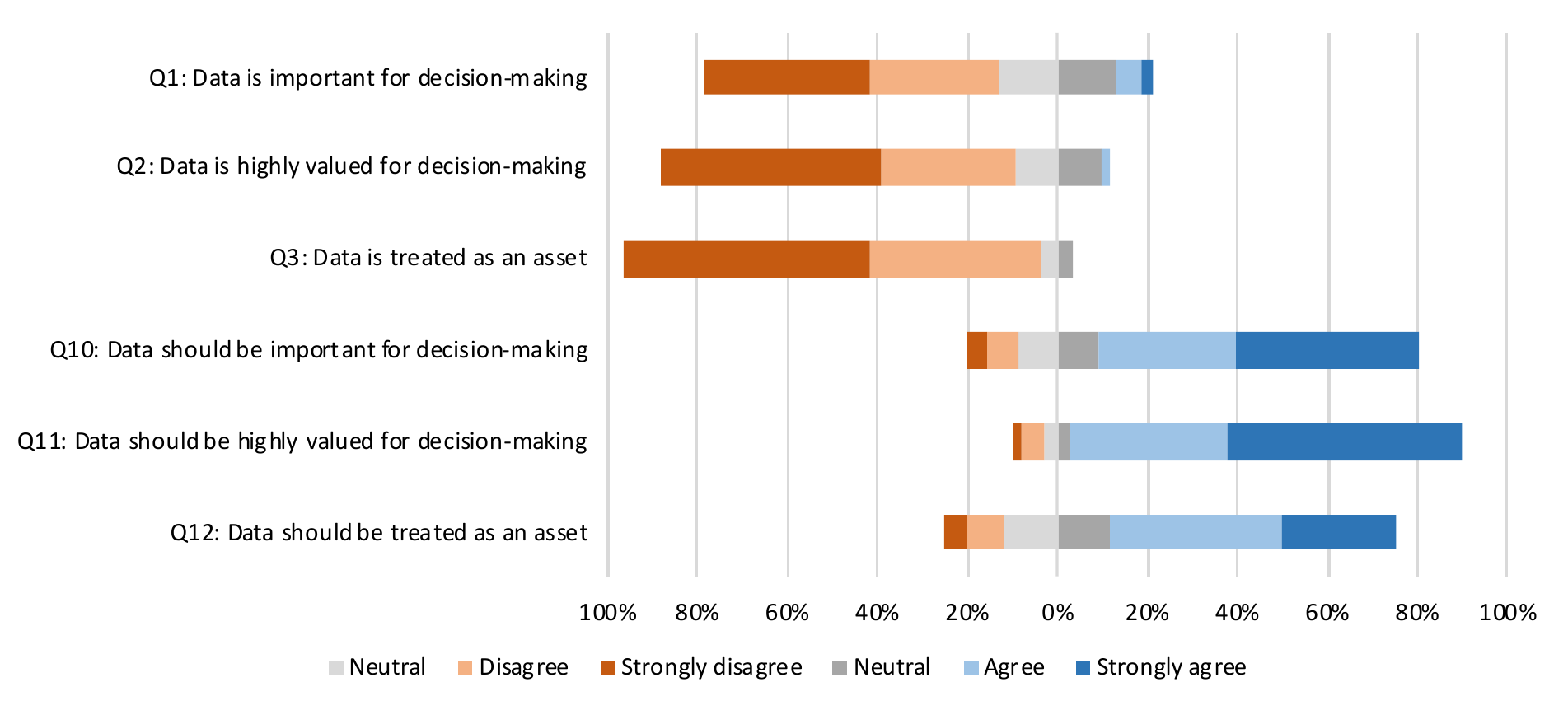}
    \caption{Respondents' view of data as part of decision making. Present (Q1--Q3) and future (Q10--Q12).}
    \label{fig:viewofdata}
\end{figure}

In analyzing Research Question 1 (RQ1), this section examines the respondents' view of data as part of decision making in ASD companies. In Figure~\ref{fig:viewofdata}, we can see the respondents' answers to each question. Each row shows the distribution of answers for that question aligned horizontally so that positive responses are to the right of the mid (zero) line while negative responses are to the left.\footnote{Note that the neutral, mid answer option (on the 5-category Likert scale) is split in half, with half of them shown in a lighter (gray) color to the left and the other half in darker (gray) color to the right of the mid (zero) line.} This makes it possible to compare the answers between different questions.

In general, looking at Figure~\ref{fig:viewofdata}, we can see that it follows the general trend identified in the statistical analysis above, i.e., respondents disagreed with the statements more in questions about the current state while agreeing more in questions about the future. For example, we see that a majority of the respondents disagreed or strongly disagreed that data is important (66\% for Q1) and highly valued (79\% for Q2) in today's decision making. However, a majority of the respondents agreed or strongly agreed that data should play an important role (71\% for Q10) and be highly valued (87\% for Q11),  when making decisions in the future. Examining if data is treated as an asset today (Q3), 93\% of the respondents disagreed or strongly disagreed, while 63\% of the respondents agreed or strongly agreed that data should be treated as an asset in the future (Q12). Although the respondents have a positive view of how data should ideally be viewed for decision making, their answers indicate this is not how it is being viewed at present in their organisations.

\subsection{Use of data in decision making (RQ2)}
\begin{figure}
    \centering
    \includegraphics[scale=0.4]{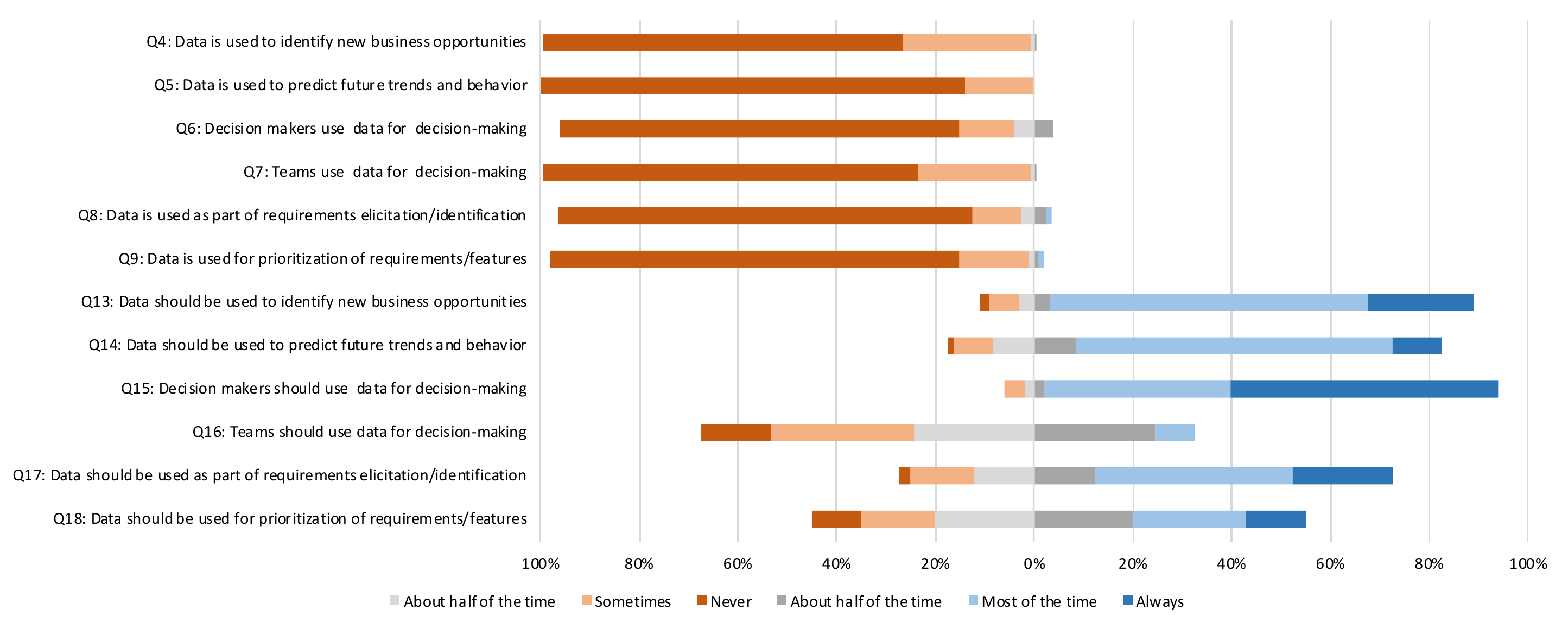}
    \caption{Use of data as part of decision making. Present (Q4--Q9) and future (Q13--Q18)}
    \label{fig:useofdata}
\end{figure}

In analyzing Research Question 2 (RQ2), this section examines to what extent data is used (present) and should be used (future) in decision making and requirements engineering in ASD companies, as illustrated in Figure~\ref{fig:useofdata}. Figure~\ref{fig:useofdata} is constructed in the same way as Figure~\ref{fig:viewofdata}, with the exception that the zero line, i.e., the neutral answer, is set to the answer `About half of the time'. In general, Figure~\ref{fig:useofdata} shows that data is seldom (never or sometimes) used in today's decision making or in Requirements Engineering (RE) (Q4--Q9 in Figure~\ref{fig:useofdata}). However, a vast majority of the respondents believe that data should be used most of the time or always in future decision making and RE (Q13--Q18 in Figure~\ref{fig:useofdata}).

Looking closely into what extent data is used in today's decision making, for all questions (Q4--Q9), more than 90\% of the respondents stated that they never or only sometimes use data in decision making and RE, where more than 73\% of the respondents stated that they never use data today. No respondent stated that they always use data. Only 1\% of the respondents stated that they use data most of the times for requirements elicitation\slash identification (Q8) and requirements prioritization (Q9). Instead of using data, the respondents explained in the free-text answer that decisions are mainly based on `gut-feeling', the decision-makers' experiences, or the value for customers. 

That is, the decisions may be subjective~\cite{Walid2016}, politically influenced~\cite{Milne2012}, and\slash or biases could be involved~\cite{Green2017}. Instead of using data when prioritizing requirements, respondents detailed that requirements are prioritized using various criteria (e.g., cost, cost\slash benefit, customer value, business value), numerical assignment, experiences, `gut-feeling', or a combination of these. This is inline with other studies on how requirements are prioritized in ASD companies today~\cite{Daneva2013}.

When asking the respondents to what extent data should be used in decision making in the future, 93\% of the respondents believe that decision makers should always, or most of the time use data for decision making (Q15), 85\% believe that data should always, or most of the time be used to identify new business opportunities (Q13), and almost 75\% believe that data should always, or most of the time be used to predict future trends and behaviours (Q14). Only 8\% of the respondents believe that (agile) teams should always, or most of the time use data for decision making (Q16), while almost half of the respondents (43\%) believe the (agile) teams should never, or only sometimes use data when making decisions. No explanation was provided by the respondents in the free-text answers for these questions.

One possible explanation may be that the respondents believe that DDDM is only useful and beneficial for high-level decisions. This is supported by the high confidence in using DDDM for identifying business opportunities (Q13) and to predict future trends and behaviours (Q14). When such high-level decisions are made, including creating product strategies, road-maps, and release plans, the respondents may believe that teams do not need DDDM when, e.g., breaking down high-level requirements to low-level ones. Another explanation may be related to today's development processes and short sprints, which may not be well suited for DDDM at the team level. 

To create and rapidly release software-intensive products in the future, it is crucial that the products are based on data and real-time feedback from the customers~\cite{Walid2016}. Thus, when moving from a subjective decision-making process, mainly based on experiences, to a DDDM process, changes in infrastructure and methodologies are needed in the development processes~\cite{Walid2016}.

For RE, 60\% of the respondents believe data should always, or most of the times be used when eliciting\slash identifying requirements in the future (Q17), while 15\% believe data should never, or only sometimes be used for requirements elicitation\slash identification. Only 35\% of the respondents believe data should always, or most of the time be used when prioritizing requirements, 25\% believe it should never, or only sometimes be used, while as many as 40\% answered that data should be used about half of the times when prioritizing requirements (Q18).

When we analyzed the data by building a simple linear model (Eq.~\ref{eq:1}) using a Bayesian approach, the results show a difference between today (`present' in Figure~\ref{fig:temp}) and the future. In Figure~\ref{fig:temp}, we see that the lower Likert scale values (e.g., answers `never' and `sometimes') are more common for Present, while the higher Likert scale values (e.g., answers `always' and `most of the time') are more common for the Future. That is, the respondents, with a high certainty, are positive to use DDDM in the future. When comparing RE related questions (Q17 and Q18) with non-RE related questions (Q13--Q16), the Bayesian model (Eq.~\ref{eq:2}) shows a difference, as shown in Figure~\ref{fig:re}. That is, although the respondents are positive to use DDDM in the future in general (as shown in Figure~\ref{fig:temp}), the respondents are more positive to use DDDM in non-RE related decisions compared to RE-related decisions. 

\textbf{Reasons for using (not using) data.} We asked the respondents what the reasons for using data in today's decision making is. According to the respondents, the main reason is that DDDM improves the decisions. One respondent explained that when data has been used as input to decision makers, the decisions have been more informed and more transparent. Another reason mentioned by the respondents was, if data is available, then we use it. 

A few respondents also gave reasons for partial data use: although the data is there and can improve decisions, it requires a lot of work to filter the data and to present the data in a way that is useful for the decision makers; thus it is only used sometimes for critical\slash important products\slash strategies.  

Looking at Table~\ref{ReasonsNotData}, we see that \textit{data is not available to us at the company} is the most common reason (82\% of the respondents). Most of the respondents who stated that data is not available, also mentioned several other reasons for not using DDDM, including \textit{too much data is available out there} (79\% of the respondents), \textit{do not know how to use the data} (73\% of the respondents), and \textit{do not know how to make the data relevant to us} (70\% of the respondents). Several of the most mentioned reasons for not using DDDM are related to the decision makers' understanding of the data (including the visualization), and how to make use of it. This confirms the findings in~\cite{Janssen2017}. In order to fully benefit from DDDM, the quality of the data is important as it is directly related to the quality of the decisions~\cite{Janssen2017}. Therefore, it is surprising that only 6\% of the respondents mentioned that data is not used in today's decision making due to the quality of the data. Either, decision making in agile is different or respondents are less aware of these important considerations.

\begin{table}
  \floatsetup{}
  \begin{floatrow}[1]
    \ttabbox%
    {\begin{tabular}{ p{8cm} | l }
      \hline
      \textbf{Reason} & \textbf{Respondents}\\ 
      \hline
      Data is not available to us at the company & 82\% \\ 
      Too much data is available out there & 79\% \\ 
      Do not know how to use the data & 73\% \\
      Do not know how to make the data relevant for us & 70\%\\
      Do not know how to link/use data in relation to decisions & 52\% \\ 
      Do not have appropriate tools & 31\% \\
      Which data should be used? & 23\% \\ 
      Cannot trust the data & 11\% \\ 
      Do not know how to access the data & 7\% \\ 
      Not sure about the quality of the data & 6\% \\
      Too many systems/tools that store the data & 4\%\\
      \hline
      \end{tabular}}
    {\caption{Reasons for not using data in decision making}
      \label{ReasonsNotData}}
\end{floatrow}
\end{table}

\subsection{How can data improve decision making (RQ3)}
We asked the respondents if they believe data could help them in making better decisions (Q19 in Table~\ref{decisionsimprovedbydata}). Eleven percent of the respondents believe data will improve their decisions (answered `yes'), while a majority (58\%) believe that data, in combination with other aspects (described below), will lead to better decisions. Close to a third (29\%) of the respondents believe data may help in making better decisions but they weren't sure (i.e., they answered `maybe'). Their stated reasons were: (1) have not used data hence do not know if it will lead to better decisions, (2) it depends on which data, the quality of the data, and who makes decisions, (3) and what kind of decisions and when the decisions are made. Only 2\% of the respondents do not believe data will help in making better decision. One respondent explained this by stating~\textit{``data can never replace my own experiences and gut-feeling''}.

\begin{table}
  \floatsetup{}
  \begin{floatrow}[1]
    \ttabbox%
    {\begin{tabular}{ p{9cm} | p{2cm} }
      \hline
      \textbf{Q19: Do you believe that you could have made better decisions if data was used as input to decision making?} & \textbf{Respondents}\\ 
      \hline
      Yes & 11\%\\
      Yes, if combined with... & 58\%\\
      Maybe & 29\%\\
      No & 2\%\\
      \hline
      \end{tabular}}
    {\caption{Respondents' views if data improves decision making}
      \label{decisionsimprovedbydata}}
\end{floatrow}
\end{table}

The respondents identified five aspects that needs to be combined with DDDM in order to make better decisions. The five aspects are: (1) own experience, (2) business value, (3) customer value, (4) input from key stakeholders, and (5) experiences from others.

In order to be able to use the full potential of DDDM and thus truly change how decisions are made in ASD, new approaches to provide and visualise constructive and understandable data (information) to the decision makers are needed. By combining understandable visualizations of data and human expertise, the future of DDDM in ASD looks promising.

\section{Conclusions}\label{Con}
There is a general trend towards data-driven decision making (DDDM), i.e., basing and driving decision making on and with data. However, there has been a lack of studies on how software practitioners view and use this and, in particular, in an agile context. In this study we thus performed a survey and collected questionnaire responses from 84 software practitioners working with agile software development.

Our main result is that the practitioners see a lot of potential for DDDM but that this potential is currently unfulfilled. While very few respondents indicated more wide-spread data-driven decision making in their current practice, a clear majority saw it as important and highly valued in the future. They were more positive to its future use for higher-level and more general decision making, fairly positive to its use for requirements elicitation and prioritization decisions, while being less positive to its future use at the team level. Multiple reasons were given for data not being used today, in particular it may not be available, be available in too large quantities, or it may not be clear how to use it, make it relevant and link it to decisions. Notably, respondents seemed less concerned about quality and trust issues around data.

Our results show that there is an unfulfilled potential for data-driven decision making in agile software development contexts. Future research should investigate this in more detail and also develop new automated data collection, analysis and visualisations techniques and methodologies that augments existing, agile decision processes by linking relevant data to specific decision contexts. 

%

\end{document}